%% file: main.tex
\documentclass[9pt,arxiv]{lapreprint}

\usepackage{pdflscape}  
\usepackage{rotating}   
\usepackage{textgreek}  
\usepackage{gensymb}    
\usepackage[misc]{ifsym} 
\usepackage{orcidlink}  
\usepackage{colortbl}   
\usepackage{tabularx}   
\usepackage{longtable}  
\usepackage{subcaption}
\usepackage{multirow}
\usepackage{snotez}     
\usepackage{soul}       
\usepackage{csquotes}   
\usepackage[textsize=scriptsize]{todonotes} 

\newcommand*{\email}[1]{\href{mailto:#1}{#1}\par}

\usepackage[
        backend=bibtex, 
    natbib=true,	
        hyperref=true,	
        alldates=year,  
    style=numeric,  
        sorting=none    
]{biblatex}         

\AtEveryBibitem{
    \clearfield{urlyear}
    \clearfield{urlmonth}
    \clearlist{language}
}

\newcommand\titlename{Revisiting the thorny issue of missing values in
 single-cell proteomics}
\title{\titlename}

\author[ \orcidlink{0000-0001-7443-5427} 1]{Christophe Vanderaa}
\author[ \orcidlink{0000-0002-1520-2268} 1 \Letter]{Laurent Gatto}

\affil[1]{Computational Biology and Bioinformatics Unit (CBIO), de
Duve Institute, UCLouvain, Belgium}

\metadata[]{\Letter\hspace{.5ex} For correspondence}{Laurent Gatto \email{laurent.gatto@uclouvain.be}}
\metadata[]{Present address}{
    Computational Biology Unit (CBIO), de Duve Institute - UCLouvain,
    Avenue Hippocrate 75, 1200 Brussels, Belgium
}

\metadata[]{Data availability}{All peptide data were retrieved from
the \texttt{scpdata} package available on
\href{https://bioconductor.org/packages/release/data/experiment/html/scpdata.html}{Bioconductor}.
The data package directly offers quantification values at precursor,
peptide and/or protein level. The code to reproduce the images
presented in the manuscript is available in the Github repository:
\hbox{\href{https://github.com/UCLouvain-CBIO/2023_scp_na}{UCLouvain-CBIO/2023\_scp\_na}}.
}

\metadata[]{Funding}{ This work was funded by a research fellowship of
    the Fonds de la Recherche Scientifique-FNRS\@.}

\metadata[]{Competing interests}{ The authors declare no competing
    interests.}

\metadata[]{Keywords}{ Single-cell, Mass spectrometry, Proteomics,
    RNA-Seq, Data analysis, Missing values, Imputation, Reproducible
    research.}

\leadauthor{Vanderaa}
\shorttitle{\titlename}

\begin{document}
\maketitle
\input{src/sections/abstract}
\input{src/sections/introduction}

\input{src/sections/impute_or_not}
\input{src/sections/challenges}

\input{src/sections/recommendations}
\input{src/sections/conclusion}
\input{src/sections/formal}
\printbibliography

\if@endfloat\clearpage\processdelayedfloats\clearpage\fi




\end{document}

%% file: src/sections/abstract.tex
\begin{abstract}

Missing values are a notable challenge when analysing mass
spectrometry-based proteomics data. While the field is still actively
debating on the best practices, the challenge increased with the
emergence of mass spectrometry-based single-cell proteomics and the
dramatic increase in missing values. A popular approach to deal with
missing values is to perform imputation.  Imputation has several
drawbacks for which alternatives exist, but currently imputation is
still a practical solution widely adopted in single-cell proteomics
data analysis. This perspective discusses the advantages and drawbacks
of imputation. We also highlight 5 main challenges linked to missing
value management in single-cell proteomics. Future developments should
aim to solve these challenges, whether it is through imputation or
data modelling. The perspective concludes with recommendations for
reporting missing values, for reporting methods that deal with missing
values and for proper encoding of missing values.

\end{abstract}

%% file: src/sections/introduction.tex
\section{Introduction} \label{intro} 

\begin{figure}
    \includegraphics[width=0.7\linewidth]{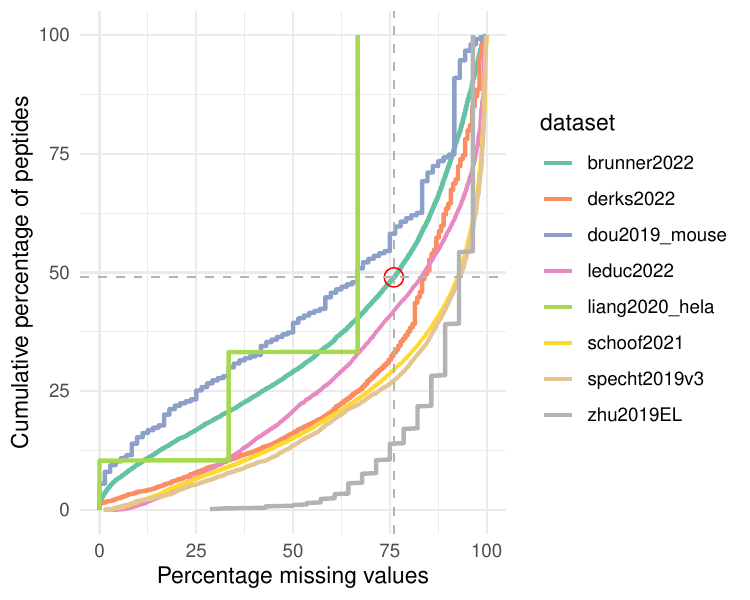}
    \caption{ \textbf{Cumulative percentage of missing
    values for different SCP datasets}. Each line (coloured by
    dataset) depicts what percentage of the peptides are retained if
    we tolerate a given percentage of missing values. For example (red
    circle), in the brunner2022 dataset, allowing at most 75 \%
    missing values would boil down to keeping only 50 \% of all
    peptides in the dataset. \label{fig:missing}}
    \figdata{ All data were retrieved from the \texttt{scpdata}
    package \citep{Vanderaa2023-qv} and systematically processed using
    the same minimal workflow. Briefly, we start from the quantified
    peptide to spectrum match or precursor data and keep only samples
    that correspond to a single cell. We then remove any feature that
    is a contaminant, matches a decoy peptide, or has an associated
    false discovery rate > 1\%. We then encode all zeros as missing
    values and filter out features that are missing for all cells.
    Finally, we compute the percentage of missing values for each
    peptide (a peptide is missing in a cell if all the corresponding
    intensities at the precursor level are missing). The percentages
    are ordered by increasing values. The percentage of peptides is
    computed as the rank of the percentage of missing values divided
    by the total number of peptides. Reference to the source data, in
    the order they appear in the legend: \citep{Brunner2022-rd,
    Derks2023-yn, Dou2019-wm, Leduc2022-cc, Liang2021-cr,
    Schoof2021-pv, Specht2021-jm, Zhu2019-ja} \label{figdata:missing}}
\end{figure}

The hurdles associated with missing values are a recurring issue in
data analysis and concern a wide range of fields and applications
\citep{Peng2023-vp}. The handling of missing values in mass
spectrometry (MS)-based proteomics is still actively debated
\citep{Kong2022-wp, Bramer2021-ml, Dabke2021-sy}. While elucidating
the best computational approaches to manage missing values is going
on, the field continues pushing the boundaries of low input
acquisitions. Recent technical advances in MS have paved the way for
MS-based single-cell proteomics (SCP) \citep{Matzinger2023-iv,
Petrosius2022-jk, Slavov2021-hf, Ctortecka2021-bh, Kelly2020-xd}, but
handling missing values in SCP data is still a clear challenge to
principled data analysis \citep{Vanderaa2021-ue}. We are confident
that the number of missing values will decrease as future technologies
will improve sensitivity \citep{Matzinger2023-iv, Derks2023-li,
Slavov2021-ab, Specht2018-hi}, but missing values will always remain
unavoidable. It is therefore important to not only think about how to
limit them but also how to deal with them during data analysis. In
this perspective, we open the discussion on how to manage missing
values in SCP data. We build our perspective on a combination of
computational advances in MS-based proteomics and in single-cell RNA
sequencing (scRNA-Seq). On the one hand, scRNA-Seq is a prolific field
with respect to computational method development dedicated to
single-cell data \citep{Zappia2021-nk}, and we foresee that many of
these efforts and tools will inspire the field of SCP. However,
handling missing values in SCP data differs from scRNA-Seq by the
presence of strong batch effects, the current lack of standardized
protocols leading to a variety of data types, and the fact the
mechanisms that generate missing values are inherently different
(count-based versus intensity-based measures). On the other hand,
MS-based proteomics already tackled many computational hurdles
associated with managing missing values \citep{Kong2022-wp}. SCP
inherits from the challenges of bulk proteomics, but these are
strongly exacerbated. While bulk proteomics datasets rarely show more
than 50 \% missing values \citep{Kong2022-wp}, over 75 \% missing
values is the rule for SCP datasets (\FIG{missing}). Moreover, strong
batch effects occur for large-scale proteomics studies and require
dedicated experimental and computational efforts to enable
reproducible results \citep{Poulos2020-tn}. SCP experiments are
large-scale studies since they require thousands of cells acquired
over at least hundreds of MS acquisitions. Finally, a major difference
between SCP and bulk proteomics is that SCP analyses can no longer
rely on replicates because every cell is unique.

Throughout the paper, we will illustrate our perspective using
published datasets retrieved from \texttt{scpdata}, an R/Bioconductor
data package. \texttt{scpdata} provides quantitative data along with
feature and sample annotations at precursor, peptide and/or protein
level. We will start by discussing the advantages and pitfalls of
imputation and highlight alternative approaches to deal with missing
values. Next, we describe the main challenges that future
computational development should aim to solve when dealing with
missing values in SCP data. Finally, we will finish by providing
recommendations on how to report the sensitivity and consistency of an
SCP experiment in terms of missing values, how to report the approach
used to deal with these missing values, and how to encode missing
values during data analysis.

%% file: src/sections/impute_or_not.tex

\section{To impute or not to impute?}


Ideally, one should avoid imputation when performing SCP data analysis
(\TABLE{imp_args}). Imputation of missing values using an unsuitable
model can lead to biased estimates and introduces false signal
\citep{Harris2023-af, Qi2022-lr, Hou2020-bw, OBrien2018-jc,
Goeminne2020-op}. Which model is suitable for which type of data is
still actively benchmarked and debated in the fields of scRNA-Seq and
bulk proteomics \citep{Hou2020-bw, Bramer2021-ml}. In scRNA-Seq,
imputation can have dramatic effects on clustering and leads to strong
artifacts in the gene expression landscape depicted by t-SNE or UMAP
projections \citep{Leote2022-ag, Hou2020-bw}. Another argument against
imputation is that imputation methods may cause over-smoothing
\citep{Qi2022-lr}, i.e. they remove biological heterogeneity because
they directly or indirectly combine observed values. This means that
any prediction is bound within the observed entries and artificially
increase the homogeneity across cells \citep{Vanderaa2021-ue,
OBrien2018-jc}. The issue of over-smoothing is of particular concern
for scRNA-Seq. Numerous methods have been developed in an attempt to
conserve the cell-to-cell relationships to identify rare cell
subpopulations \citep{Qi2022-lr, Hou2020-bw, Leote2022-ag}. Another
important limitation is that imputation is an estimation process with
inherent uncertainty. Replacing missing values with a single point
estimate ignores the variability associated with the estimation
process \citep{Hou2020-bw, OBrien2018-jc}. For instance, highly
parameterized imputation methods may lead to over-fitting that is
characterized by high variance in the prediction \citep{Tran2022-hb}.
Multiple imputation approaches attempt to tackle this issue by
computing multiple predictions for each entry. These predictions are
then summarized, typically by averaging, leading to estimates that are
more robust than using a single prediction \citep{Gong2018-ze,
Giai_Gianetto2020-ct}. Furthermore, the variance associated with each
imputation and across imputations can be combined and used for
downstream analysis \citep{Chion2022-pe}, although this substantially
complicates the analytical approach. The performance of multiple
imputation however still relies on the performance of the underlying
imputation method used \citep{Lazar2016-zl} and does not solve
systematic bias. Finally, imputation can mask contrasts that are
inestimable \citep{OBrien2018-jc}. Protein quantities are derived from
peptide intensities. However, peptides are often quantified in a
limited fraction of the cells. When two groups of cells share no
quantified peptides and there is no third group of cells that
expresses these peptides to serve as a reference, the contrast is
considered inestimable. Imputation will artificially fill the gaps and
mask these inestimable contrasts, leading to highly biased estimates
\citep{OBrien2018-jc}. We also emphasize that aggregation of peptides
intensities to protein intensities in the presence of missing values
leads to implicit imputation \citep{Lazar2016-zl}. For example,
aggregating peptides using the median expression implicitly assumes
that the missing values equal the estimated median. Another example is
the aggregation of peptides using the sum. This will implicitly impute
missing values by zero. Implicit imputation suffers from the same
drawbacks mentioned above but occurs unnoticed and hence undocumented.

\input{src/tables/imputation_arguments.tex}


An alternative approach to imputation is to apply models that account
for missing values. In proteomics for example, the selection model for
proteomics (SMP, \cite{OBrien2018-jc}) or the hurdle model
\citep{Goeminne2020-op} are models that contain a component for
modelling peptide abundance and another to model peptide missing rates
while taking into account sample and batch effects. While both models
provide a compelling solution to missing data for proteomics, it only
applies to label-free experiments and can only be applied to a single
task, differential analysis. Single-cell downstream analyses also
typically include clustering, dimension reduction or trajectory
inference \citep{Amezquita2020-bf, Luecken2019-fc}. Moreover, SMP is
based on a Bayesian framework optimized through Gibb's sampling, and
the hurdle model is fit using a mixed model framework. Hence, these
methods will unlikely scale to large SCP datasets. Another solution is
to rely on dimension reduction output, such as principal component
analysis (PCA), to perform the downstream analysis tasks. Alternating
least squares algorithms such as NIPALS or regularized expectation
maximization (EM-)PCA algorithms such as missMDA are two approaches
that perform PCA in the presence of missing values
\citep{Josse2016-qd}. However, the former may suffer from numerical
instability and the latter is computationally expensive and does not
scale for large SCP datasets. Moreover, both approaches assume the
missing value pattern is random, which does not hold for both
proteomics and scRNA-Seq data
\citep{Giai_Gianetto2020-ct, OBrien2018-jc, Tran2022-hb}, and especially
for SCP \citep{Vanderaa2021-ue}. Different methods have been developed
for scRNA-Seq data to tackle these limitations. For instance, the
ZINB-WaVE and scVI are modelling approaches that perform dimension
reduction while accounting for missing values, batch effects and
normalization \citep{Risso2018-sa, Lopez2018-ye}. These models also go
beyond dimension reduction. ZINB-WaVE estimates the probability that
missing entries follow a negative binomial distribution. The
probabilities can be used to leverage bulk differential analysis tools
\citep{Van_den_Berge2018-rp}. scVI directly integrates an inference
step in its approach enabling the assessment of statistical
differences between subpopulations of interest. 


Although imputation should ideally be avoided, many SCP analyses
include imputation in their workflow \citep{Vanderaa2023-qv}. The
reasons are mainly practical (\TABLE{imp_args}). First, many
downstream methods cannot directly deal with missing values. The tools
that implement them either return an error when missing values are
present or will output unusable results because the missing values are
propagated to every entry of the data. Using a suitable imputation
method to predict missing values unlocks these methods. Another reason
is that it can be included as a standalone step in a workflow.
Although it is wrong to assume that imputation is independent of other
processing steps such as normalization or batch correction
\citep{Vanderaa2021-ue}, it dramatically facilitates the assessment of
its performance \citep{Hou2020-bw, Tian2019-cm}. This assessment is
much harder for methods such as ZINB-WaVE or scVI where the quality of
the outcome will depend on the different components of the model
without the possibility to pinpoint performance bottlenecks. Also,
different data require different imputation methods. For instance,
\citet{Hou2020-bw} and \citet{Kong2022-wp} provide decision trees for
which imputation method to select given data characteristics for
scRNA-Seq and proteomics, respectively. Hence, allowing to easily plug
a different imputation method in a workflow is easier and quicker
than developing a new model tailored for each data type and analysis
task at hand. Finally, imputation methods can easily be transferred
and reassessed across fields. For instance, K-nearest neighbour (KNN)
imputation is an imputation method used in a wide range of application
and fields \citep{Hastie2001-zs} and has been rapidly adopted by the
SCP field that still lacks dedicated models to handle missing values
\citep{Vanderaa2023-qv}.


In conclusion, we recommend avoiding imputation, but in practice, this
can be difficult to achieve. Suitable methods to handle missing values
are still to be developed for SCP. While imputation leads to
inaccuracies, it still leverages useful information from complex data
\citep{Hou2020-bw}. In the following section, we list five main
challenges that should be considered when developing imputation
methods or models that handle missing values for SCP data. 

%% file: src/tables/imputation_arguments.tex
\begin{table}[!ht]
    \caption{
        \textbf{Arguments against and for imputation.}
        \label{tab:imp_args}}
    \begin{tabularx}{\linewidth}{p{0.11\textwidth} p{0.85\textwidth}}
        \toprule
        \multicolumn{2}{c}{\textbf{To impute or not to impute?}} \\
        \midrule
        \textbf{No} & 
        \begin{itemize}
            \item Imputation introduces bias
            \item Imputation artificial increases homogeneity across
            samples
            \item Imputation omits the variance associated to the
            estimation process
            \item Imputation masks inestimable contrasts
        \end{itemize} \\
        \textbf{Alternatives} &
        \hspace{1.5em}Use methods that can account for missing values:
        \begin{itemize}
            \item Models dedicated to differential analysis: SMP, 
            hurdle model
            \item Dimension reduction as start point for downstream
            analyses: NIPALS, missMDA
            \item All-in-one models: ZINB-WaVE, scVI (only scRNA-Seq)
        \end{itemize} \\        
        \textbf{Yes} &
        \begin{itemize}
            \item Imputation unlocks many computational tools that
            break upon missing values
            \item Imputation is easier to develop and assess as a
            separate step
            \item Imputation is easily transferable across fields
        \end{itemize} \\
        \bottomrule
    \end{tabularx}
\end{table}

%% file: src/sections/challenges.tex

\section{Challenges in managing missing values in SCP data}

\subsection{Challenge 1: High proportion of missing values}


The first challenge is to cope with high proportions of missing values
(\TABLE{challenges}). A typical SCP dataset can contain on average
between 50 and 90~\% missing values. In the context of scRNA-Seq, most
imputation methods exhibit a decreased performance for transcripts
with high proportions of missing values \citep{Leote2022-ag}. This is
expected since, on the one hand, fewer data are available to predict
the missing value and, on the other hand, more bias is introduced by
performing more estimations. Up to now, the methods used to impute SCP
data were taken from the proteomics field \citep{Vanderaa2023-qv}, but
their suitability to single-cell data needs to be evaluated. The
performance of imputation methods on bulk data has been evaluated on
datasets with, relatively, low numbers of samples and missing value
proportions of about 20 to 50 \% \citep{Kong2022-wp, Bramer2021-ml,
Lazar2016-zl, Webb-Robertson2015-uq}. Another question raised by the
high proportion of missing values in SCP data is how to perform
peptide or protein filtering. Bulk proteomics pipelines often remove
highly missing peptides, usually allowing at most 90~\% of missing
values \citep{Kong2022-wp}. Depending on the SCP datasets, this filter
can lead to removing between 50 and 80 ~\% of all quantified peptides
(\FIG{missing}). One challenge is therefore to leverage useful
information from these highly missing peptides. Another challenge is
that the performance of imputation methods depends on the proportions
of missing values. \citet{Kong2022-wp} suggest different methods for
imputing missing values based on the proportion of missing values.
This is also observed for scRNA-Seq. In this context,
\citet{Leote2022-ag} conclude that ``there is no universally
best-performing method that outperforms the others in all cases''.
Therefore, they propose an ensemble approach where the best imputation
method is selected for each transcript separately using a
cross-validation approach. We expect this relationship between the
accuracy of imputation (or data modelling) and the proportion of
missing values to hold for SCP, and ensemble approaches may offer an
interesting framework to tackle this challenge.

\input{src/tables/challenges_summary.tex}

\subsection{Challenge 2: Different types of data}


The second challenge is the diversity of data acquisition protocols
and instruments to perform SCP (\TABLE{challenges}). For instance, we
can classify protocols as label-free (LFQ) or as multiplexed (TMT,
mTRAQ). LFQ experiments acquire one cell per MS run, while multiplexed
experiments acquire from 3 to 18 cells per MS run. These two protocols
lead to different correlation structures and batch effects in the
data. Furthermore, samples processed by either of these two protocols
can be acquired using data-dependent acquisition (DDA) or
data-independent acquisition (DIA). Also, DDA can be targeted towards
specific peptides known to be present in a sample by using a
prioritized inclusion list for MS2 selection \citep{Huffman2023-cj,
Leduc2022-cc}. A thorough description of the different technologies
for acquiring SCP data is provided by others \citep{Petrosius2022-jk}.
These protocols, in combination with the instruments to run them,
generate different types of data and lead to different patterns of
missing values, implying that different imputation assumptions and
algorithms might be required. This is not specific to SCP. In bulk
proteomics, benchmarking efforts and literature reviews focus on one
type of data. For instance, \citet{Kong2022-wp} focus on DDA LFQ data,
\citet{Bramer2021-ml} focus on DDA TMT data and \citet{Dabke2021-sy}
focused on LFQ DIA data. Hence, the development and validation of SCP
imputation methods should be dataset-specific.

\subsection{Challenge 3: Cell to cell heterogeneity}


The imputation method should ideally account for cell-to-cell
heterogeneity (\TABLE{challenges}). Each single cell is a unique
sample, hence no technical replication is possible
\citep{Gatto2023-kk}. Solutions to this challenge have been explored
in the field scRNA-Seq. For instance, MAGIC \citep{Van_Dijk2018-ua}
improves upon sample-wise KNN imputation by taking into account the
similarity between cells (we will discuss the meaning of sample-wise
KNN in the next section). When imputing a value with sample-wise KNN,
we first search a discrete set of K cells that are closely related to
the cell to impute. The average of these neighbouring cells provides
the imputation estimate. The number of neighbours K will influence the
imputation results. We have previously shown that KNN
imputation on SCP data can lead to exceedingly smooth profiles
for highly missing peptides or proteins \citep{Vanderaa2021-ue}.
Instead of searching for a discrete set of neighbours, MAGIC computes
the probability of transitioning from one cell to another and uses the
probabilities as weights to estimate the missing value for a given
cell. This approach takes into account that each cell is different and
may contribute differently to the imputation process. However, MAGIC
borrows information from cell-cell relationships, hence it can
potentially influence these relationships for downstream analysis,
such as clustering of trajectory inference. A similar imputation
approach is to predict missing values using gene-gene relationships,
as implemented in the scISR or the Network algorithms
\citep{Qi2022-lr, Leote2022-ag}. The drawback of these methods is that
building a reliable gene network is difficult and either relies on
fully observed genes \citep{Qi2022-lr} or on external bulk data
\citep{Leote2022-ag}.

\subsection{Challenge 4: Strong batch effects}


The fourth challenge is the presence of strong batch effects in SCP
data (\TABLE{challenges}). SCP experiments contain hundreds to
thousands of cells that must be analysed over many MS runs. Slight
differences in experimental conditions across MS acquisitions
influence the identification and quantification of peptides
\citep{OBrien2018-jc}. We also observe a strong influence of batch
effects on peptide quantification and missing rate
\citep{Vanderaa2021-ue}. An imputation method for SCP should account
for batch effects as well. The ProNorm method was designed to account
for these effects in the context of large bulk proteomics datasets
acquired using a DIA approach \citep{Poulos2020-tn}. ProNorm first
removes batch effects using the RUV-III-C model and then performs
imputation by sampling from a normal distribution centred around the
average of the batch-corrected replicates. The method has however
several drawbacks when applied to single-cell data. First, it relies
on the presence of replicates. As stated above, we can no longer rely
on replicates with single-cell data. Second, the model also relies on
negative controls, i.e. peptides that are known to be constant across
cells. The presence of such negative controls is again a strong
assumption in the context of single cells given the presence of
cell-to-cell heterogeneity. Finally, the RUV-III-C model is
specifically designed to remove any variance that is not associated
with biological factors and/or experimental conditions of interest. In
other words, RUV-III-C removes any variability that is not explained
by the experimental design. This property is not desirable for
single-cell analyses as the main objective is to leverage unexplained
variance, for instance, to discover rare cell states or continuous
differentiation processes that are blurred by bulk approaches. This
method could be adapted to SCP similarly to how ZINB-WaVE adapted the
RUV model to scRNA-Seq data \citep{Risso2018-sa}.

\subsection{Challenge 5: Different causes of missing values}

\begin{figure}
    \includegraphics[width=0.7\linewidth]{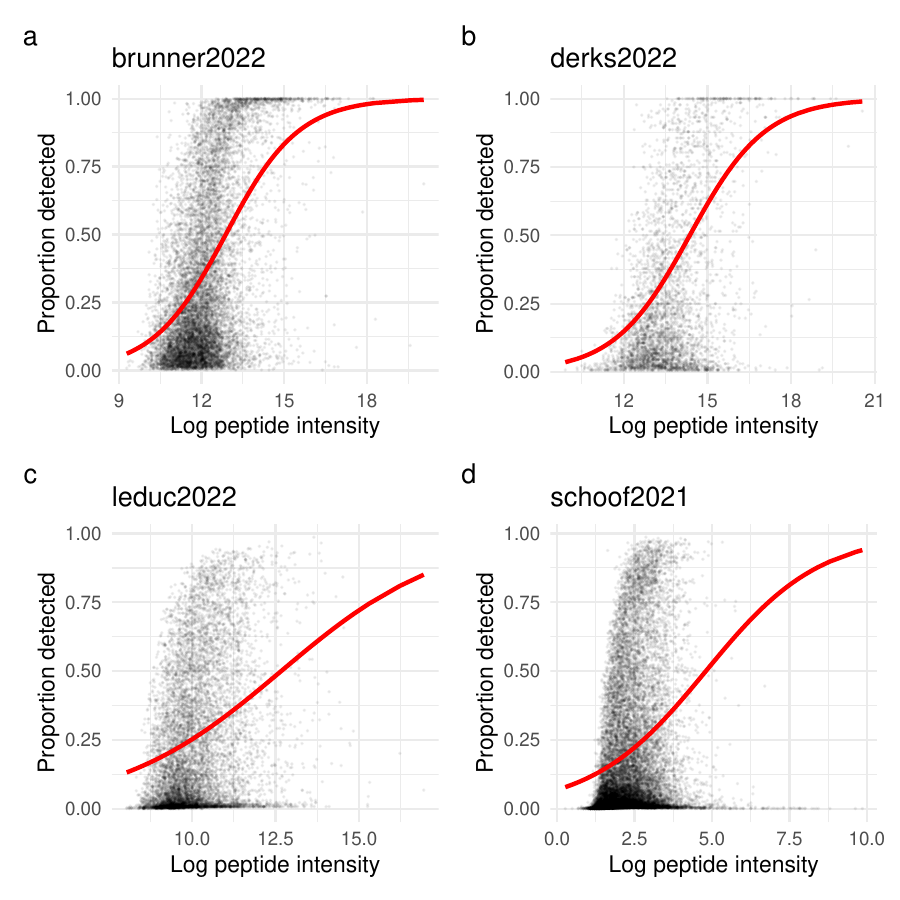}
    \caption{\textbf{Relationship between detection
    probability and observed peptide intensity.} The relationships are
    shown for 4 representative SCP datasets: brunner2022 (\textbf{a},
    LF-DIA), derks2022 (\textbf{b}, mTRAQ-DIA), leduc2022 (\textbf{c},
    TMT-DDA) and schoof2021 (\textbf{d}, TMT-DDA). Each point
    represents a peptide in the dataset. Peptide intensities are
    estimated as the average of the log intensities across all cells
    (ignoring missing values) as suggested by \citet{Li2023-rg}. The
    proportion of detection is the number of samples where the peptide
    is observed divided by the total number of samples. The red line
    represents the fitted detection probability curve and was
    estimated using the \texttt{prodDP} R package \citep{Li2023-rg}.
    The poor fit indicates the relationship between detection
    probability and observed peptide intensity is weak for SCP data.
    \label{fig:MNAR} \figdata{All data were retrieved from the
    \texttt{scpdata} package \citep{Vanderaa2023-qv} and
    systematically processed using the same minimal workflow. Briefly,
    we start from the quantified peptide to spectrum match or
    precursor data and keep only samples that correspond to a single
    cell. We then remove any feature that is a contaminant, matches a
    decoy peptide, or has an associated false discovery rate > 1\%. We
    then encode all zeros as missing values and filter out features
    that are missing for all cells. Data were further log-transformed.
    When multiple precursors map to the same peptide, the intensities
    are aggregated using the median. The logistic regression curves
    were fit using the \texttt{stats::glm()} function in R.
    \label{figdata:MNAR}} }
\end{figure}

The fifth challenge is to account for the different mechanisms that
generate missing values (\TABLE{challenges}). There are 2 types of
mechanisms: biological missingness and technical missingness. A
peptide or protein is biologically missing from a cell if that cell
does not express the feature. A peptide or protein is technically
missing from a cell if the instrument or the computational workflow
was not able to detect and quantify the feature. Biological and
technical missingness bear different information and distinguishing
the two leads to biologically meaningful insights
\citep{Goeminne2020-op}. However, this distinction is difficult to
assess accurately and formulate mathematically. Instead, the
proteomics field often classifies missing values as missing completely
at random (MCAR) and missing not at random (MNAR) \citep{Kong2022-wp,
Lazar2016-zl}. MCAR means that the process that generated the missing
value is completely random and unrelated to the true underlying value.
Conversely, MNAR means that the missing value probability is dependent
on the underlying value. In other words, a peptide may be MNAR because
its corresponding protein is not present in a cell or at levels that
fall below the MS2 threshold for selection (DDA) or limit of detection
of the instrument. A peptide may be MCAR because of instrumental
artefacts (e.g. slight variation in ionization efficiency) or
computational limitations (e.g. poor identification score). Different
imputation algorithms are required to correctly impute MNAR or MCAR
values \citep{Kong2022-wp, Lazar2016-zl}. For instance, the sampling
from a down-shifted normal distribution imputation approach is
recommended for missing values that exhibit MNAR properties, while a
dimension reduction-based imputation is recommended when the missing
values are MCAR \citep{Kong2022-wp}. However, SCP datasets and more
generally large proteomics datasets exhibit both MCAR and MNAR in
unknown proportions \citep{Vanderaa2021-ue, Bramer2021-ml,
Webb-Robertson2015-uq}. In line with the second challenge, the
prevalence of MNAR and MCAR will depend on the type of datasets, but
it will also depend on the nature of the peptide and protein analysed
\citep{Li2023-rg, Giai_Gianetto2020-ct, OBrien2018-jc}. This
complicates the development and assessment of imputation methods.
A solution is to estimate the
proportion of MNAR and MCAR using probability theory, as already
explored for bulk proteomics \citep{Li2023-rg, Ahlmann-Eltze2020-zg,
Giai_Gianetto2020-ct}. \citet{Giai_Gianetto2020-ct} suggest a multiple
imputation framework where estimated MCAR/MNAR probabilities are used
to stochastically select an MCAR- or MNAR-devoted imputation
algorithm. \citet{Ahlmann-Eltze2020-zg}, and more recently
\citet{Li2023-rg}, developed a statistical approach where detection
probabilities are estimated using a logistic regression model to infer
the bias caused by ignoring MNAR. Both models assume a strong
relationship between detection probability and measured intensities
and were developed for label-free data. However, multiplexing is
foreseen to play an important role in achieving the high sample
throughput required for single-cell applications \citep{Derks2023-li,
Specht2018-hi}. We do not find empirical evidence of such a
relationship in SCP data, especially for multiplexed datasets
(\FIG{MNAR}). Future efforts should consider cell-to-cell
heterogeneity and batch effects when deciphering the missing data
mechanisms in SCP data.

We addressed the challenges separately, but in practice, new imputation
methods or modelling approaches for SCP data should address these
simultaneously. Therefore, we expect that more complex algorithms than
those currently used for SCP data processing will be required. These
challenges are currently overwhelming and hard to tackle, but we are
optimistic. We foresee that future advances in the technologies and
search engines will reduce the number of missing values generated in
single cells \citep{Derks2023-li, Boekweg2023-qb, Slavov2021-ab}.
Hence, tackling these challenges will become easier with time as the
quality of the data to impute will improve.

%% file: src/tables/challenges_summary.tex
\begin{table}[!ht]
    \caption{ 
        \textbf{Summary of the main challenges in managing
        missing values in SCP data.} \label{tab:challenges}
    }
    \begin{tabularx}{\linewidth}{p{0.28\textwidth} p{0.65\textwidth}}
        \toprule
        \textbf{Challenges} & \\
        \midrule
        High proportion of missing values & SCP data contain between
        50 and 90~\% missing values. This leads to decreased
        imputation performance and requires to reassess the thresholds
        used for filtering peptides or proteins based on the
        proportion of missing values.\\
        Different types of data & SCP data can be acquired using DIA
        or DDA approaches, LFQ or multiplexed protocols, using
        orbitraps or time of flight. The diversity of technologies
        leads to different missing value patterns.\\
        Cell to cell heterogeneity & Every cell is unique, hence no
        technical replicates are available. Several methods already
        tackle this challenge when imputing scRNA-Seq data: MAGIC,
        scISR, Network.\\
        Strong batch effects & SCP data is acquired through hundreds
        of MS runs, leading to inevitable batch effects both on the
        identification and the quantification. The RUV-III-C method is
        a promising approach, provided adaptation to SCP data.\\
        Different causes of missing values & Values may be missing 
        because of MNAR or MCAR mechanisms. Most methods to deal with 
        missing values are tailored towards one or the other 
        mechanism. A potential solution is to apply cross-validated 
        ensemble approaches, as suggested for scRNA-Seq, or 
        relying on probability estimations, as as suggested for bulk
        proteomics. \\
        \bottomrule
    \end{tabularx}
\end{table}

%% file: src/sections/recommendations.tex

\section{Recommendations for managing missing values in SCP data}

In this section, we extend the initial recommendations recently
published for performing, benchmarking, and reporting SCP experiments
\citep{Gatto2023-kk}, focusing on reporting missing values and the
methods to manage them. We will also discuss how to encode missing
values in SCP data.

\subsection{Reporting sensitivity and consistency}

\begin{figure}
    \begin{fullwidth}
        \includegraphics[width=0.9\linewidth]{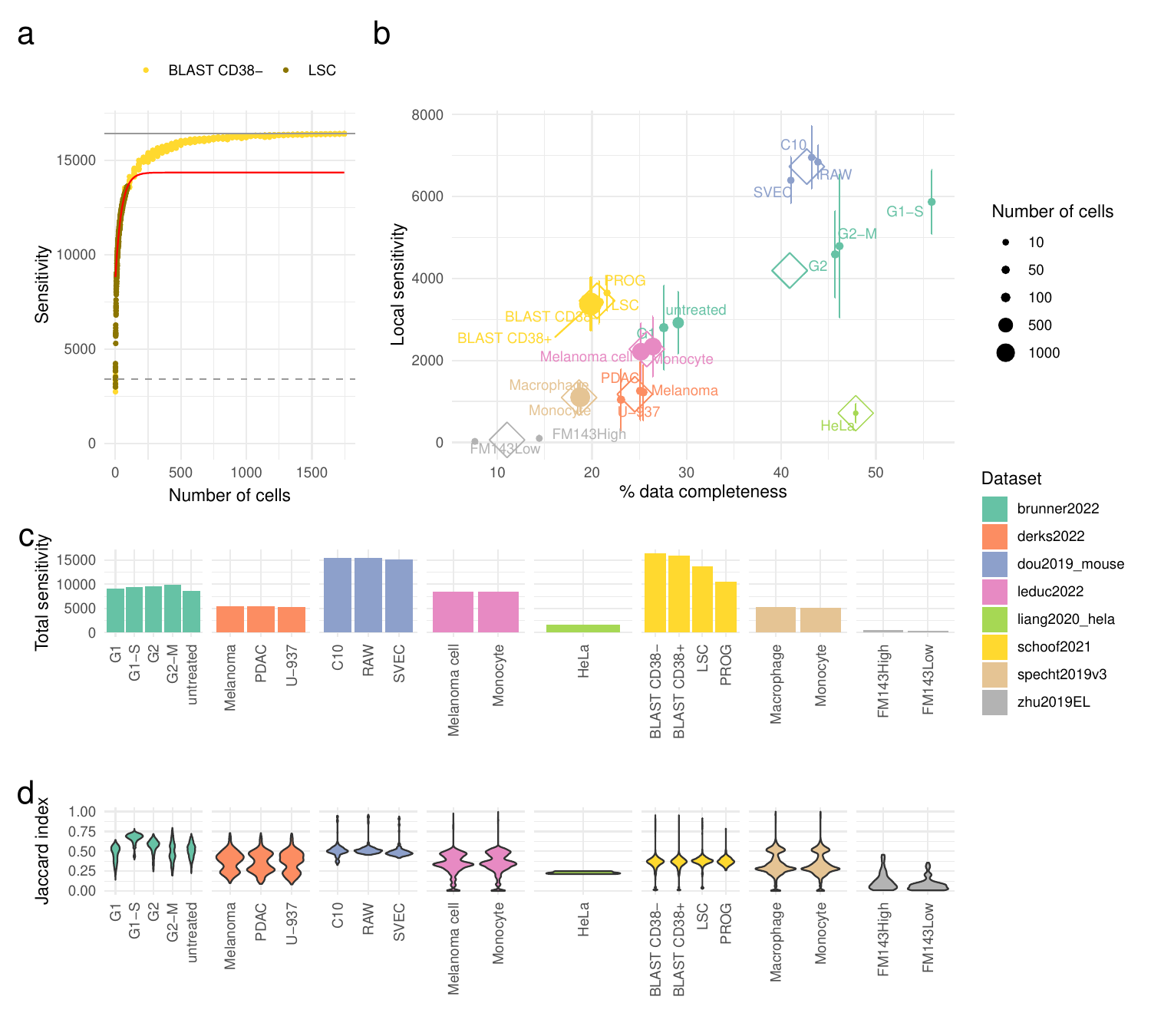}
        \caption{\textbf{Reporting sensitivity. a.} Cumulative
        sensitivity curve showing the total number of distinct
        peptides as we sample an increasing number of CD38- blast
        cells (yellow) or leukaemic stem cells (LCS, brown) from the
        Schoof et al. dataset. The dashed line represents the local
        sensitivity, that is the average sensitivity of a single cell.
        The solid grey line is the total sensitivity. The solid red
        line shows the fit of the asymptotic regression model for the
        LSC data, with extrapolation. \textbf{b.} The median local
        sensitivity with respect to data completeness. Each point is a
        cell type, coloured according to the corresponding dataset,
        and sized according to the number of cells acquired. The
        vertical lines represent the standard deviation. Empty
        diamonds indicate the dataset average. \textbf{c.} Total
        number of distinct peptides found in various cell types
        grouped by dataset (colour). \textbf{d.} Distribution of the
        pairwise Jaccard index within each cell type for each dataset.
        \label{fig:sensitivity}}
        \figdata{All data were retrieved from the \texttt{scpdata}
        package \citep{Vanderaa2023-qv} and systematically processed
        using the same minimal workflow. Briefly, we start from the
        quantified peptide to spectrum match or precursor data and
        keep only samples that correspond to a single cell. We then
        remove any feature that is a contaminant, matches a decoy
        peptide, or has an associated false discovery rate > 1\%. We
        then encode all zeros as missing values and filter out
        features that are missing for all cells. We then build the
        identification matrix where rows correspond to peptides and
        columns to single cells, and entries are 1 if the peptide was
        found in the corresponding cell and 0 otherwise. The matrix is
        then split based on the cell type. Metrics are computed as
        follows. The number of cells is the number of columns of the
        matrix. The local sensitivity is the sum by column. The total
        sensitivity is the number of rows and the data completeness is
        the average of the matrix. Jaccard indexes are computed for
        each pair of columns as the size of the union of peptides
        divided by the size of the intersection. The cumulative
        sensitivity plot was constructed from the identification
        matrix as described in the body of the text. References to the
        source data, in the order they appear in the legend: 
        \citep{Brunner2022-rd, Derks2023-yn, Dou2019-wm, Leduc2022-cc,
        Liang2021-cr, Schoof2021-pv, Specht2021-jm, Zhu2019-ja}
        \label{figdata:sensitivity}}
    \end{fullwidth}
\end{figure}

Missing values and sensitivity are two sides of the same coin. If a
technology had perfect sensitivity, no missing data would be generated
other than biologically missing values. We here define sensitivity as
the number of distinct peptide or protein identifications. To
facilitate the discussion, we will not distinguish between sensitivity
at the peptide or protein level because they show the same trends, and
we will refer to features instead. Sensitivity is commonly reported as
the number of features per cell \citep{Derks2023-yn, Webber2022-vk,
Woo2022-rt, Leduc2022-cc, Brunner2022-rd}. We call this metric the
\textit{local sensitivity} (\FIG{sensitivity}a, dashed line, and b).
Local sensitivity is computed for each cell, hence this metric should
be summarised with the mean and standard deviation. While
technological developments aim to maximize local sensitivity, it
should be interpreted with care. Solely looking at the local
sensitivity does not provide the full picture of the number of
features that can be detected in an experiment. Every cell is unique,
and every run is subject to technical variations that influence
feature identification. So, we do not expect to find all detectable
features by looking at individual cells. Therefore, we recommend 
also reporting sensitivity as the total number of distinct features
identified across the whole experiment. We call this the \textit{total
sensitivity} (\FIG{sensitivity}a, solid line, c). It allows a more
accurate estimation of how many features a given experiment can
potentially detect in low-amount samples. Total sensitivity has also
associated caveats. When reporting total sensitivity, we assume that
enough cells were acquired to capture all detectable features. This is
the case for the CD38 negative blast cells acquired by Schoof et al.
\citep{Schoof2021-pv}. By including thousands of cells in their
experiment, the authors were able to reach a saturation of the set of
identified peptides (\FIG{sensitivity}a). If they had limited their
analysis to less than 500 blast cells, they would have underestimated
the total sensitivity. This is the case for leukaemic stem cells,
which are scarcer than blast cells. About a hundred stem cells do not
allow to reach the total sensitivity for that cell type. We expect
that reaching total sensitivity will be difficult for rare cell
populations and will offer a solution later in the section.

Although local and total sensitivity are useful metrics when reported
together, they provide no information about the consistency of
identification. Consistency of identification relates to the
similarity between sets of features identified across different cells
\citep{Boekweg2023-qb}. Consistency of identification between cells
has previously been reported using Venn diagrams, or upset plots
\citep{Derks2023-yn, Woo2022-rt, Webber2022-vk, Vanderaa2021-ue,
Liang2021-cr, Dou2019-wm}. These approaches nevertheless limit the
comparison to tens of single cells, otherwise, these visualizations
become difficult to interpret. We recommend instead to report the
\textit{data completeness}. Data completeness is the proportion of
observed values across the dataset. It relies on the fact that the
lower the consistency between cells, the more missing values will be
incorporated in the dataset, hence the lower the data completeness. As
soon as a cell with new features is added to a dataset, these new
features will introduce missing values for the other cells and
therefore decrease the data completeness. But again, reporting only
the data completeness is not sufficient. It provides no information
about sensitivity. For instance, the Liang et al. dataset showed a
relatively high data completeness (hence high consistency), but at the
expense of a limited number of features (low total sensitivity,
\FIG{sensitivity}b, \FIG{sensitivity}c). Another drawback is that data
completeness depends on the accurate estimation of the total
sensitivity, hence the need to acquire a sufficient number of cells. The
Liang et al. dataset is composed of only three single cells and hence
underestimates the total sensitivity. This explains the apparently
high but biased data completeness.

In conclusion, we suggest reporting at least the local sensitivity,
the total sensitivity, the data completeness and the number
of cells included in the dataset. No useful assessment of sensitivity
and consistency is possible without reporting these four metrics
together. They are simple to implement and do not require complex data
analysis. These metrics represent an objective way to compare
different parameters in an SCP experiment that may affect sensitivity
and consistency, such as sample preparation protocols, MS settings or
peptide identification software. Only one of these parameters should
be varied if the objective is to compare performance. Therefore,
\FIG{sensitivity}c cannot be used to suggest the superior performance
of a protocol over the others because the datasets contain different
cell types, they were acquired with different instruments, and they
were processed with different search engines. Still, we can see a
correlation between the local sensitivity and the data completeness,
indicating that increased sensitivity allows retrieving more of the
same features across cells. We also recommend computing these metrics
for the full dataset and for each cell type separately (defined in the
experimental design or after data analysis). Different cell types
express different sets of peptides in different quantities which will
inevitably influence the sensitivity. Finally, we focus the discussion
solely on reporting the sensitivity of identification. We refer to the
initial recommendations for a thorough discussion about the assessment
of the quantitative data \citep{Gatto2023-kk}, and we refer to the
recent review by \citet{Boekweg2023-qb} for recommendations about the
computational challenges for peptide and protein quantification in SCP
data. 

We next extend upon the four metrics with two more advanced concepts
that may be harder to implement. 

First, we recommend testing whether the number of cells is sufficient
to accurately assess the total sensitivity. We suggest reporting a
cumulative sensitivity plot (\FIG{sensitivity}a), where the
sensitivity is reported as a function of the number of cells. To build
this graph, we randomly sample an increasing number of cells from the
dataset and compute the sensitivity for each pool. We suggest
performing several draws with the same number of cells to include an
estimation of the variance associated with the stochastic selection
process. The sensitivity is expected to plateau at the total
sensitivity as we include more cells (\FIG{sensitivity}a). This can be
modelled using an asymptotic regression model. In case the total
sensitivity is not achieved, the total sensitivity may be extrapolated
using the fitted model asymptote (\FIG{sensitivity}a, red line).

Second, inspired by the quality assessment by \citet{Derks2023-yn}, we
also recommend reporting the Jaccard index between all possible
pairs of cells within a cell type (\FIG{sensitivity}d). The Jaccard
index measures the overlap between the sets of features identified
from 2 cells. Contrarily to the data completeness, the Jaccard index
no longer depends on the total sensitivity. However, computing the
Jaccard index between all pairs of cells is more elaborate and is best
performed using a programming language such as R or Python. 

\subsection{Reporting methods for missing values}

\begin{figure}
    \includegraphics[width=\linewidth]{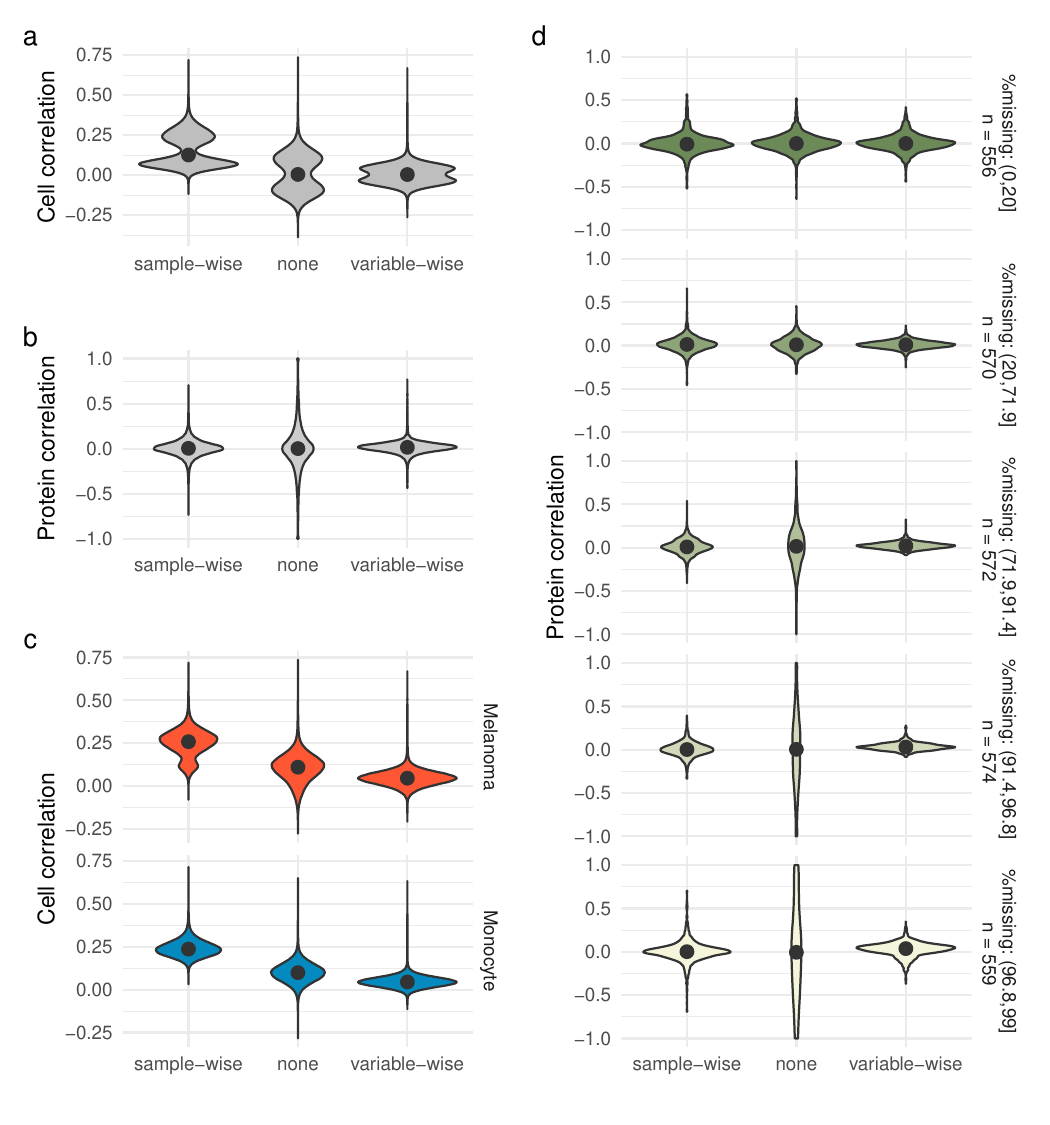}
    \caption{\textbf{Sample-wise and variable-wise KNN are not the
    same. a.} Distribution of Pearson correlations computed for each
    pair of cells. Data were either not imputed (none), sample-wise
    KNN imputed or variable-wise KNN imputed. Note that the observed
    values before imputation are not modified after imputation. Points
    highlight the median of the distribution.\textbf{b.} Same as a.,
    but for protein correlation. \textbf{c.} The correlations in a.\
    are shown for pairs of cells within a cell type, that is melanoma
    (red) or monocyte (blue).\textbf{d.} The correlations in b.\ are
    separated into 5 groups of proteins depending on the percentage of
    missing values. Groups are ordered from lowly missing (dark green)
    to highly missing (pale yellow). The data are taken from Leduc et
    al. \citep{Leduc2022-cc}. \label{fig:knn_imputation}} \figdata{The
    nPOP data \citep{Leduc2022-cc} were retrieved from the
    \texttt{scpdata} package \citep{Vanderaa2021-ue}. We took the
    protein data processed before imputation \citep{Vanderaa2023-qv}.
    To avoid complicating the interpretation with confounding effects,
    we remove batch effect using the \texttt{removeBatchEffect}
    algorithm from the \texttt{limma} R package (version 3.55.1). Note
    that we do not recommend this approach for SCP data analysis (c.f.
    fourth challenge). Next, we impute the protein data using the
    \texttt{impute} function from the \texttt{QFeatures} package
    (version 1.9.2). We used the KNN algorithm with \texttt{K = 15},
    ensuring that every entry is imputed with KNN, regardless of the
    proportion of missing values (\texttt{rowmax = 1, colmax = 1}) and
    disabling computational shortcuts (\texttt{maxp = Inf}). Pairwise
    correlation per cell and per protein were then computed. When two
    cells have the same type, their correlation is annotated with that
    cell type. The proteins were binned into 5 categories depending on
    their percentage missing values. When two proteins are included in
    the same bin, their correlation is flagged with that bin.
    \label{figdata:knn}}
\end{figure}

The minimal requirement for reporting an imputation method or a
missing data model approach is to provide the name of the algorithm
used, the name of the software that implements it and the version of
that software. We stress that these three elements are the bare
minimum for a meaningful and useful method report. Reporting the
software version is critical, because different versions of the same
software may add, remove or update functionality, hence possibly
changing the output. Next to reporting the version, we have seen
Method sections that solely report a piece of
software or an algorithm \citep{Vanderaa2023-qv}. This leads to
ambiguity since software often implement several algorithms, and an
algorithm is often implemented by multiple pieces of software. For
instance, \texttt{VIM} (R), \texttt{impute} (R) and \texttt{sklearn}
(python) offer functionality to perform KNN imputation. These three
pieces of software offer different parameters (and default values),
perform different optimizations, solve edge cases differently, and
have different assumptions about the data. We experienced the impact
of these differences when replicating published SCP data analyses
\citep{Vanderaa2021-ue}. \citet{Specht2021-jm} performed KNN
imputation using their own implemented in R. We observed significant
numerical differences when comparing their processed data tables to
those generated by our R/Bioconductor package \texttt{scp}
\citep{Vanderaa2021-ue}. After inspection, we were able to identify
the origin of the problem. The reimplementation by Specht performs
sample-wise KNN (i.e. infers missing values from closely related
cells), while our \texttt{scp} package relies on the
\texttt{impute.knn} function from the R package \texttt{impute} that
performs variable-wise KNN (i.e. infers missing values from closely
related peptides or proteins). \FIG{knn_imputation} shows the impact
of both KNN approaches on cell and protein correlations. Because
sample-wise KNN borrows information from other cells, it will on
average increase correlations between cells (\FIG{knn_imputation}a and
c). Conversely, variable-wise KNN will remove correlation between
cells because it focuses on the relationships between features. We
also observe a small impact on protein correlations
(\FIG{knn_imputation}b). We observe a reduction in the protein
correlations after both sample-wise and feature-wise KNN imputation,
but we also notice a subtle but systematic shift towards higher
correlation when imputing with sample-wise KNN. Since the proportion
of missing values vary drastically across proteins, we binned the
protein correlation based on their percentage of missing values
(\FIG{knn_imputation}d). KNN imputation, both sample-wise and
variable-wise, has little impact on correlation distribution when
proteins have less than 20 \% missing values (\FIG{knn_imputation}d,
top). This is expected since fewer predictions are required. As more
observations are missing, the protein correlations before imputation
become inaccurate. Sample-wise KNN imputation of highly missing
proteins restores the correlation distributions centred around zero,
as seen when proteins are less missing. Variable-wise KNN leads to
correlation distributions that are sharper and progressively shifted
towards higher correlations as the percentage of missing values
increases. 

Sample-wise and variable-wise KNN rely on different assumptions
leading to numerical differences that will propagate through the data
processing workflow. In theory, sample-wise KNN leads to a violation
of the sample independence assumption on which most statistical
methods rely. Similarly, the same violation occurs for proteins
when combining variable-wise KNN and statistical methods to infer
regulatory protein networks. However, how these violations impact the
statistical outcome in practice is still to be explored. 

To conclude, we strongly recommend reporting the missing value
management approach by mentioning the algorithm, the name of the
software and its version together. However, we stress that the best
practice is to provide the code (e.g. scripts, Rmarkdown files or
Jupyter notebooks) that generates the published results along with the
computational environment containing all the software tools in the
state that they were used to run the code \citep{Gatto2023-kk}. 

\subsection{Encoding missing values}
\label{section_encoding}

Unobserved entries in scRNA-Seq data are encoded as zeros. This is
because RNA-Seq is a count-based approach where zero is a meaningful
readout. Proteomics experiments, however, are based on methods that
measure electrical current intensities. In the absence of any signal,
MS instruments still record noise levels as low values, but zeros are
not expected to occur. Any zero present in SCP data is generated
computationally, either because the intensities below a given
threshold were trimmed to zero or because quantifications for
unobserved peptides or proteins were filled with zeros when joining
multiple samples together. Since these computationally derived zeros
are not meaningful, we recommend replacing them with a dedicated data
type, such as the \texttt{NA} value in R or the \texttt{numpy.nan}
value in Python. Larger datasets will require sparse data
representations that are crucial for managing and handling the data
with limited computational resources. These data representations, such
as those implemented in the \texttt{Matrix} R package or the
\texttt{scipy.sparse} Python module assume the missing entries are
zeros. Future large-scale analyses might make use of sparse
representations to support efficient data storage. Zeros should then
be considered placeholders that are disregarded during data processing
and downstream analyses. Otherwise, this would boil down to an
implicit zero imputation that is likely ill-suited for SCP data
\citep{Kong2022-wp, Lazar2016-zl}. We suggest future computational
tools for SCP data to handle missing values and zero values with great
care.

%% file: src/sections/conclusion.tex
\section{Conclusion} \label{conclusion}

Handling missing values in SCP can be an intimidating task. We prefer
to avoid imputing missing values, but unfortunately no alternatives
have yet been suggested that are easily applicable to SCP. Although
imputation methods can and are applied to SCP data, their suitability
still needs to be demonstrated. Computational benchmarking efforts are
required to pinpoint the limitations of these imputation methods on
SCP data, but also to assess the practical significance of the
limitations associated with imputation compared to data modelling.

We advise that future computational developments consider 5 main
challenges when dealing with missing values. First, the proportion of
missing values is high. Most datasets have over 70 \% missing entries
at the peptide level. At these rates of missing values, several
imputation methods fail to provide accurate predictions
\citep{Kong2022-wp}. Second, there is a diversity of experimental
approaches to SCP. Each approach has its strengths and weaknesses and
leads to different missing data patterns. Different algorithms might
be necessary to deal with each pattern, and we do not expect that an
algorithm that performs best for one type of data will perform best
for all types of data. Therefore, future computational innovation must
clearly state on which type of data they were tested and are expected
to be best suited. Third, the imputation methods should take into
account that every single cell is unique and that replicates are not
available. This will require new methods to evaluate the sweet spot
between over-fitting and over-smoothing. Fourth, SCP inevitably
imposes large-scale experiments. Therefore, batch effects are and will
be present in SCP data \citep{Vanderaa2021-ue}. We cannot imagine a
successful method that handles missing values without handling batch
effects as well. Finally, missing values may be deterministic and/or
stochastic, and distinguishing the two mechanisms may avoid bias in
the data analysis. However, to what extent the bias impacts downstream
analyses still requires thorough investigation for SCP data.

We provide a few recommendations targeted toward SCP practitioners.
First, we recommend reporting at least four metrics when reporting the
sensitivity of an experiment. The local sensitivity provides
information about the number of features that are detected in a single
cell. Total sensitivity provides information about the number of
features that are potentially detectable in the experiment. The data
completeness provides the global agreement between the sets of
features identified across cells. The number of cells acquired during
the experiment provides essential information because total
sensitivity and data consistency require enough cells to be correctly
estimated. We recommend not interpreting these metrics when less than
100 cells are acquired, and suggest using a cumulative sensitivity
plot to assess whether the number of cells allows for an accurate
estimation of the total sensitivity. Reporting the pairwise Jaccard
index provides also useful information about data consistency. We
provide functions to compute the suggested metrics in the \texttt{scp}
R/Bioconductor package \citep{Vanderaa2021-ue}. A tutorial showing how
to compute these metrics is available on the package's webpage.
\footnote{\url{https://uclouvain-cbio.github.io/scp/articles/reporting_missing_values.html}}
We also recommend providing at least 3 pieces of information when
reporting the approach used to deal with missing values: the name of
the algorithm, the name of the software, and the version of the
software. We also strongly recommend providing the code and the
software environment to reproduce any published results. This
recommendation is not new and was already presented previously
\citep{Gatto2023-kk}. We however provide a comprehensive example
showing that small variations in applying an imputation method can
have a significant impact on the processed data and downstream
analyses. Finally, we raise concerns regarding how missing values are
encoded. We recommend encoding missing values using an appropriate
data type rather than a numeric value (e.g. \texttt{NA} for R and
\texttt{numpy.nan} for Python).

The field of SCP is rapidly expanding \citep{noauthor_2023-mq},
but it craves for adequate solutions to deal with missing data. This
represents a great opportunity for computational biologists and
bioinformaticians that want to join the exciting advances of SCP.

%% file: src/sections/formal.tex
\subsection{Acknowledgment}

This preprint was created using the LaPreprint template
(\url{https://github.com/roaldarbol/lapreprint}) by Mikkel Roald-Arb\o
l \textsuperscript{\orcidlink{0000-0002-9998-0058}}.

We thank Lieven Clement, Davide Risso, Charlotte Soneson and the
members of the CBIO lab for useful discussions about scRNA-Seq and 
proteomics data imputation.

\subsection{Author contributions}

Conceptualization: L.G.; 
Methodology: C.V., L.G.; 
Software: C.V., L.G.;
Formal analysis: C.V.; 
Writing - original draft: C.V., L.G.; 
Writing - review \& editing: L.G.; 
Visualization: C.V.; 
Supervision: L.G.; 
Project administration: L.G.; 
Funding acquisition: L.G. 